\documentclass[twocolumn,showpacs,preprintnumbers,amsmath,amssymb]{revtex4}

\usepackage{graphicx}
\usepackage{dcolumn}
\usepackage{bm}
\begin{document}


\title{Crystal growth and in-plane optical properties of Tl$_2$Ba$_2$Ca$_{n-1}$Cu$_n$O$_x$ (n=1,2,3) superconductors}

\author{Y. C. Ma}
\author{N.L. Wang}
\altaffiliation[Email: ]{nlwang@aphy.iphy.ac.cn}
\affiliation{%
Beijing National Laboratory for Condensed Matter Physics,
Institute of Physics, Chinese Academy of Sciences, Beijing 100080,
People's Republic of China
}%

\date{\today}

\begin{abstract}
Single crystals of thallium-based cuprates with the general
formula Tl$_{2}$Ba$_{2}$Ca$_{n-1}$Cu$_{n}$O$_{x}$(n=1,2,3) have
been grown by the flux method. The superconducting transition
temperatures determined by the ac magnetic susceptibility are 92
K, 109 K, and 119 K for n=1,2,3 respectively. X-ray diffraction
measurements and EDX compositional analysis were described. We
measured in-plane optical reflectance from room temperature down
to 10 K, placing emphasis on Tl-2223. The reflectance roughly has
a linear-frequency dependence above superconducting transition
temperature, but displays a pronounced knee structure together
with a dip-like feature at higher frequency below T$_c$.
Correspondingly, the ratio of the reflectances below and above
T$_{c}$ displays a maximum and a minimum near those feature
frequencies. In particular, those features in Tl2223 appear at
higher energy scale than Tl2212, and Tl2201. The optical data are
analyzed in terms of spectral function. We discussed the physical
consequences of the data in terms of both clean and dirty limit.
\end{abstract}

\maketitle

\section{INTRODUCTION}

Since the discovery of the high temperature superconductors(HTSC)
in the 1980's, many experiments have been done to disclose the
microscopic origin of high temperature superconductivity. For
example, inelastic neutron scattering revealed a peculiar magnetic
resonance mode at an energy of 41 meV in two-dimensional
reciprocal lattice position ($\pi$,$\pi$) for optimally doped
YBa$_{2}$Cu$_{3}$O$_{6+x}$\cite{Norman,Mook}. The angle resolved
photoemission spectroscopy (ARPES) has indicated a kink in the
band dispersion as well as an anisotropic gap (the d-wave
behavior) for some high-T$_c$ cuprates.\cite{Norman,Ding,Loeser}.
These results are very helpful for understanding the
superconductivity mechanism in the cuprates. Naturally, one may
ask whether these behaviors are generic in HTSC materials. But
there are technical reasons why those experimental probes have not
been used with success on all the families of HTSCs. For example,
inelastic neutron scattering experiment requires large-size single
crystals and had mostly been done on YBCO-123 and
LSCO-214\cite{Norman,Mook,Cheong}, although some groups have tried
to work on Tl$_{2}$Ba$_{2}$CuO$_{6+\delta}$ by synthesizing about
300 relatively large (0.5 to 3mm$^3$) single crystals and
co-aligning them in a mosaic of total volume 0.11cm$^3$\cite{He},
as well as on Bi$_{2}$Sr$_{2}$CaCu$_{2}$O$_{8+\delta}$ with a
comparatively large single crystal of volume
10$\times$5$\times$1.2mm$^3$ and mosaicity (that is, the angular
spread of the crystallographic axes),
$\sim$1${^\circ}$\cite{Fong}. Tunnelling spectroscopy and ARPES
are surface-sensitive probes. They are mainly applied to Bi-based
cuprate systems as they could be cleaved easily in a vacuum along
the Bi-O planes, yielding a high-quality virgin
surface\cite{Norman}. However, technique like optical spectroscopy
places relatively less demands on the sample size and surface, and
have consequently been applied with success to a larger number of
high temperature superconducting systems. At present, optical
probes cover a wide frequency and temperature range, which have
provided much useful information about charge excitations and
dynamics of cuprate superconductors.

In the whole HTSC families, thallium-based cuprates have provided
a large series of superconductors including single Tl-O layered
compounds with the number of CuO$_2$ layers from 1 to 5 and double
Tl-O layered compounds Tl$_{2}$Ba$_{2}$Ca$_{n-1}$Cu$_{n}$O$_{x}$
with the CuO$_2$ layers n from 1 to 3 (hereafter abbreviated as
Tl-2201, Tl-2212, Tl-2223,
respectively)\cite{Parkin1,Ihara,Sheng1,Toshihiro,Maignan,Kondoh,Parkin2,Merrien}.
The Tl-based cuprates offer a good opportunity to investigate the
physical properties of systems with different number of CuO$_2$
layers in a unit cell, and therefore different T$_c$ at the
optimal doping. This will help to understand the mechanism of high
temperature superconductivity. Another advantage of Tl-based
cuprates is that, all the double Tl-O layer based cuprates have
higher $T_{c}$s than the double Bi-O layer based cuprates with the
same structure. Even for the single CuO$_2$ layered compound
Tl-2201, its T$_c$ can reach 90 K. Consequently, the difference
between the superconducting state and the normal state can be
easily probed. Nevertheless, despite those advantages, much less
spectroscopic studies have been done on the Tl-based cuprates.
This is mainly due to the difficulty of obtaining high quality
single crystal samples. One has to overcome the problem of
avoiding the poisonous thallium volatility and formation of
intergrowth defects during crystal growth\cite{Maignan}. Infrared
studies have been done on crystals of single-layered
Tl-2201\cite{Puchkov1,Puchkov2}, and thin-films of double-layered
Tl-2212\cite{Wang}, but there is still no report, to the best of
our knowledge, on triple-layered Tl-2223 and any other Tl-based
superconductors.

We have recently successfully grown single crystals of
Tl$_{2}$Ba$_{2}$Ca$_{n-1}$Cu$_{n}$O$_{x}$ compounds with n=1,2,
and 3 by the flux method. In this study, we first describe the
crystal growth procedure and characterization of the as grown
crystals by the ac magnetic susceptibility, X-ray diffraction
measurements and EDX compositional analysis. Then, we report the
measurement of the in-plane infrared reflectance. As optical data
for optimally doped Tl-2201 and Tl-2212 are available in
literature, we place emphasis on the spectra collected on
triple-layered Tl-2223 crystals and their comparisons with data of
Tl-2212 and Tl-2201. Very similar to the Tl-2201 and Tl-2212, we
observe significant change of spectra after entering
superconducting state, which was related to the combination of a
Boson mode and a gap. We analyzed the spectra in terms of spectral
function. A linear scaling is established between the energy of
the characteristic feature and the superconducting transition
temperature for different systems at optimal doping.

\section{CRYSTAL GROWTH AND CHARACTERISTICS}

Single crystals were grown by the flux method from raw materials
with nominal compositions as listed in Table I. The mixed oxide
BaCaCuO powders were synthesized prior to the crystal growth. For
example, we made the BaCaCuO-223x powder for Tl-2212, and
BaCaCuO-266x for Tl-2223. Then, Tl$_{2}$O$_{3}$ powder was added
into the mixed powders to get a stoichiometry as in Table I. The
ratio of the oxide powders is essential for the crystal phase
growth.

In the growth experiment, the powders of amount $\thicksim$15
grams were ground in an agate mortar, then placed in an alumina
crucible and covered with two Al$_{2}$O$_{3}$ lids for decreasing
the volatility of Tl$_{2}$O$_{3}$. The crucible was put in a tube
furnace with flowing oxygen, and heated up to a high temperature
for several hours. The temperature was set to be different for
growing Tl-2201, Tl-2212, and Tl-2223 systems as shown in Fig. 1.
The furnace was cooled down slowly at a rate of about
10${^\circ}$C per hour to certain temperature (depending on the
system), then down to room temperature naturally. The oxygen
flowed all the time during the sintering process. The starting
temperature and the cooling rate seriously affect the size of a
as-grown crystal. It also affects substantially the
superconducting transition temperature of Tl-2201 phase.

\begin{table}
\caption{\label{tab:table2}Nominal compositions in starting
materials for crystal growth and transition temperatures of grown
crystals Tl-Ba-Ca-Cu-O\cite{Toshihiro}:}
\begin{tabular}{cccccc}
\hline%
    Sample &Starting material &$T_{c}$     &Size            &Phase\\
      No.   &Tl/Ba/Ca/Cu      &K           & mm$^2$                \\
\hline
      A     & 2/2/0/1       & 92  & 1.2$\times$1.5 &Tl-2201 \\
      B     & 2/2/2/3       & 109 & 2.0$\times$1.8 &Tl-2212 \\
      C     & 2/2/6/6       & 119 & 1.2$\times$0.9 &Tl-2223 \\
\hline
\end{tabular}
\end{table}

\begin{figure}
\includegraphics[width=7.0cm,height=5.0cm]{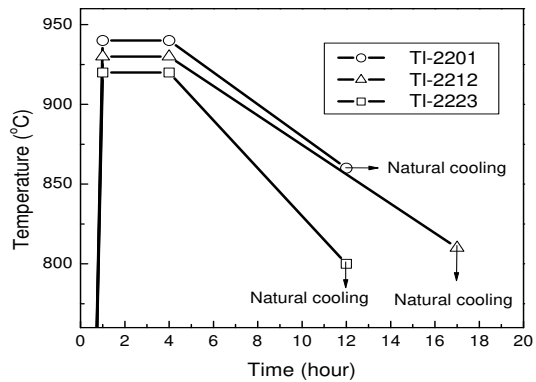}
\caption{\label{fig:epsart} The growth process for Tl-2201(a),
Tl-2212(b) and Tl-2223 (c) single crystals. }
\end{figure}



As the crucible had been cooled down to room temperature, the
mm-sized crystals of Tl-2201, Tl-2212, and Tl-2223 single crystals
have formed. Some shinny, usually free-standing crystals are found
at the place near the cavities formed in the melt. The SEM image
of a typical Tl-2223 crystal is shown in Fig. 2.

\begin{figure}
\includegraphics[width=7cm,height=5cm]{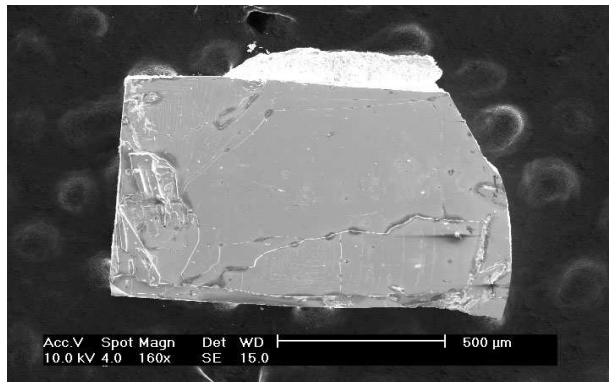}
\caption{\label{fig:epsart} SEM images of as grown crystals
Tl-2223.}
\end{figure}

According to X-ray diffraction measurements shown in Fig. 3,
nearly all the (0,0,$l$) peaks can be seen clearly, furthermore,
no other phase can be seen in the diffraction pattern, and every
sample has a good c-axial orientation which is perpendicular to
the sample natural growth face. The half widths of midpoint of the
X-ray diffraction peaks are typically 0.07$^{\circ}$,
0.09$^{\circ}$ and 0.08$^{\circ}$ for Tl-2201, Tl-2212, and
Tl-2223 single crystals, respectively. Clearly the intergrowth
defects of the different members have been depressed by the
sintering process. The lattice constants, c, of the three type of
crystals are 23.21{\AA}, 29.28{\AA} and 35.55{\AA} for Tl-2201,
Tl-2212, Tl-2223 phases, respectively, as there is a CuCaO$_{2}$
difference in each unit cell for the three phases.

\begin{figure}
\includegraphics[width=7cm,height=9.0cm]{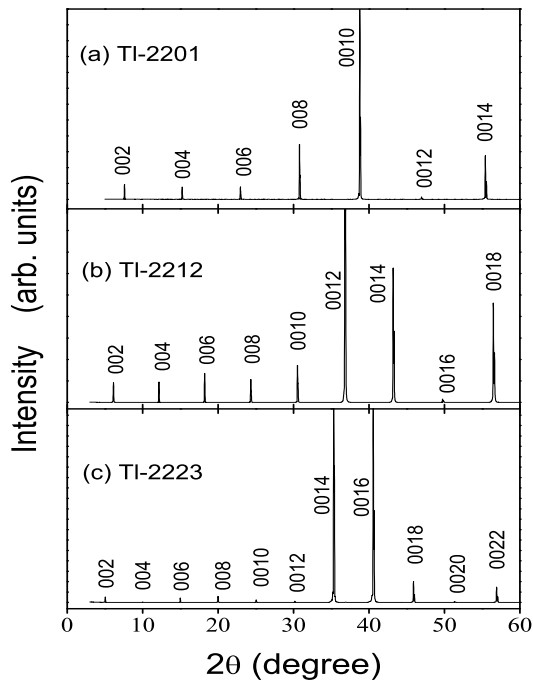}
\caption{\label{fig:epsart} The X-ray diffraction patterns up to
60 degree for (a)Tl-2201,(b)Tl-2212 and (c)Tl-2223.}
\end{figure}

The superconducting transition temperature T$_c$ was determined by
ac susceptibility measurement. T$_c$ increases in the three phases
as the CuO$_2$ layers in a unit cell increase from 1 to 3. The
T$_c$ for as-grown Tl-2212 and Tl-2223 crystals are always close
to 109 K and 119 K, respectively, while the T$_c$ for as-grown
Tl-2201 crystals ranges from about 70 K to 92 K. Fig. 4 shows the
temperature dependence of the ac susceptibility for as-grown
Tl-2201, Tl2212, and Tl-2223 single crystals.

\begin{figure}
\includegraphics[width=7.0cm,height=5.0cm]{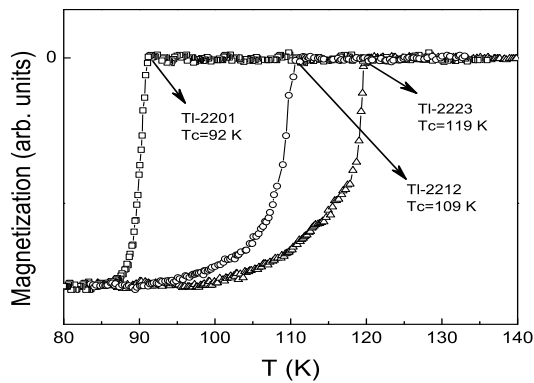}
\caption{\label{fig:epsart} Temperature dependence of ac
susceptibility for Tl-2201, Tl-2212 and Tl-2223 single crystals.
The maximum magnetization in each sample has been normalized.}
\end{figure}

The compositions of the crystals were examined by energy
dispersive X-ray micro analysis(EDX). The quantitative
measurements were performed at several points on the surface of
the crystal, which show a good uniformity of composition on the
crystal surface. Typical EDX spectra obtained from each phase of
the crystal are shown in Fig. 5. As expected form the ideal
chemical formula of each phase, the peak ratios of
CaK$\alpha{/}$BaL$\alpha$ and CuK$\alpha$${/}$BaL$\alpha$ in the
EDX spectra increased with the number of Cu-O layers in Tl
compounds. Note that there is no Ca in the Tl-2201 phase, this is
because we did not mix the CaO before this phase has been formed.
Other peaks are of the Tl, Ba, Ca, Cu character spectrum. All
these results are consistent with the X-ray diffraction
experiments.

\begin{figure}
\includegraphics[width=7cm,height=9.0cm]{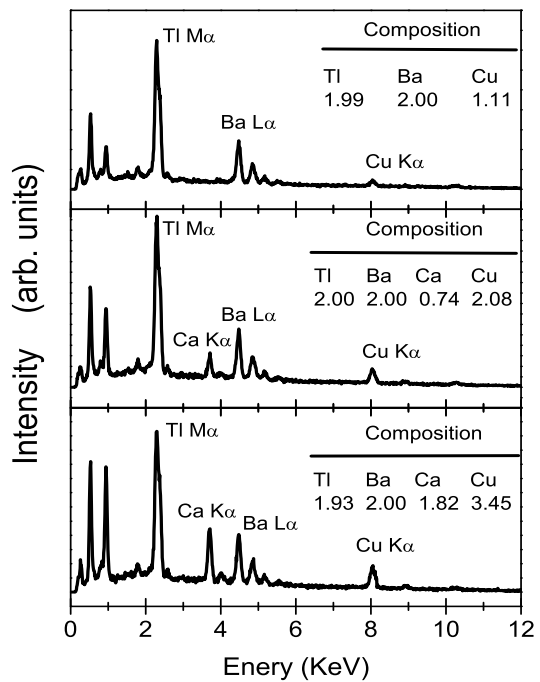}
\caption{\label{fig:epsart} EDX spectra and analyzed compositions
for Tl-2201(a),Tl-2212(b) and Tl-2223 (c) phase.  The compositions
are calculated on the assumption that the molar ratio of Ba is the
ideal value of two.}
\end{figure}

\section{Optical Properties }

Optical spectroscopy measurement can yield rich information about
charge dynamics of a system. Intensive optical studies have been
done on La-, Y-, and Bi-based cuprate systems, comparatively, much
less optical investigations have been done on Tl-based systems. As
we mentioned in the introduction, infrared studies on the Tl-based
systems have been performed only on single-layer and double-layer
compounds near optimal doping. Therefore, in this work, we shall
mainly focus on the data of Tl-2223 crystals, as well as their
comparisons with data of Tl-2201 and Tl-2212.

The reflectance measurements from 100 to 22000 cm$^{-1}$ were
carried out on a Bruker 66v/S spectrometer. An in-situ overcoating
technique was used for the experiment\cite{Homes}. The optical
conductivity spectra were derived from the Kramers-Kronig
transformation. Hagen-Rubens relation was assumed for the
low-frequency extrapolation. At high frequency side, a constant
extrapolation was adopted up to 100,000 cm$^{-1}$, then a
R($\omega$)$\thicksim$$\omega$$^{-4}$ was used. A comparison of
the in-plane reflectances of Tl-2201, Tl-2212 and Tl-2223 single
crystal at room temperature is shown in Fig. 6. Clearly as the
CuO$_2$ layers of a unit cell goes from 1 to 3, the reflectance
edge goes to higher energy. The observation indicates that the
carrier density increases with increasing the number of CuO$_2$
layers in a unit cell.

\begin{figure}
\includegraphics[width=7.0cm,height=5.0cm]{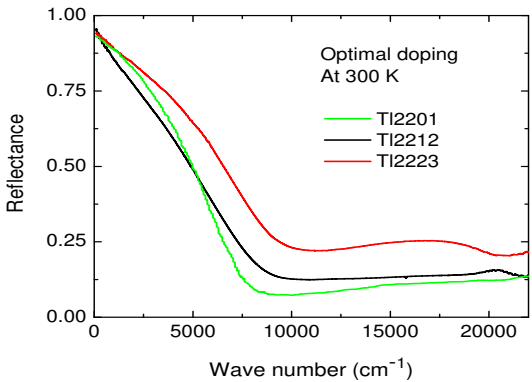}
\caption{\label{fig:epsart} (Color online) The reflectance data of
Tl-2201, Tl-2212 and Tl-2223 crystals up to 22000 cm$^{-1}$ at
room temperature. }
\end{figure}

Fig. 7 (a) and (b) show the reflectance R($\omega$) and
conductivity ${\sigma}$$_{1}$(${\omega}$) spectra for a Tl-2223
crystal at different temperatures. In the high temperature, the
Tl-2223 reflectance spectra show roughly linear frequency
dependence, a behavior common to all hole-doped high-T$_c$
cuprates. The linear frequency dependence of $R(\omega$) implies a
linear variation of the inverse lifetime of carriers with
frequency, which could be well described by a marginal Fermi
liquid theory.\cite{Hwang} With decreasing temperature, the
low-$\omega$ $R(\omega$) increases, indicating a metallic response
of the sample. At 10 K in the superconducting state, R($\omega$)
shows a clear knee structure at around 600 cm$^{-1}$. Above this
frequency, the reflectance drops fast and becomes lower than the
normal-state values at around 1000 cm$^{-1}$. R($\omega$) recovers
the linear-frequency dependence at higher frequency about 1600
cm$^{-1}$. Similar but slightly weak behaviors were seen at 90 K,
which is close to $T_{c}$.

\begin{figure}
\includegraphics[width=7.0cm,height=9.0cm]{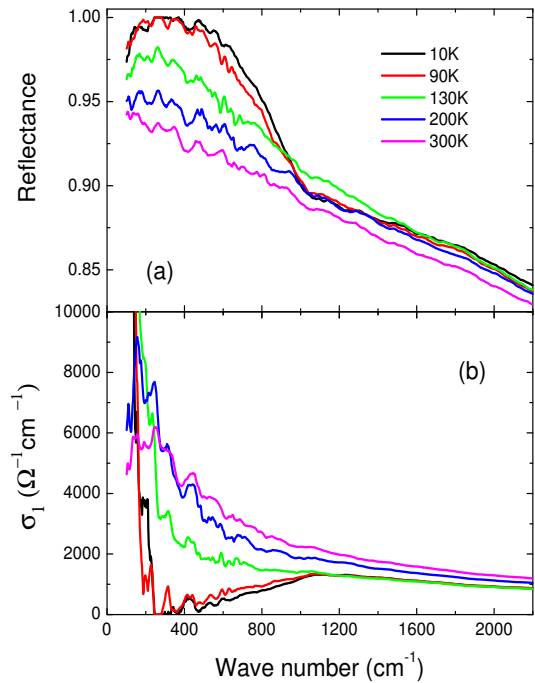}
\caption{\label{fig:epsart} (Color online) ab-plane optical data
of the optimally doped Tl-2223 single crystal with $T_{c}$=119K.
(a) The temperature dependent reflectance and (b) the
temperature-dependent ${\sigma}_{1}$(${\omega}$). }
\end{figure}

The temperature- and frequency-dependent optical responses of
Tl-2201 and Tl-2212 crystals are very similar to those of Tl-2223
sample. For comparison we show in Fig. 8 the conductivity spectra
for optimally doped Tl-2201, Tl-2212 and Tl-2223 at 300 K and 10
K, respectively. Qualitatively, the $\sigma_1(\omega)$ spectra
exhibit the same characteristic in both normal and superconducting
states for the three phases. The conductivity spectrum has a
Drude-like shape in the normal state, but becomes suppressed below
a certain frequency in the superconductiing state. However, it is
found that the energy scale below which the $\sigma_1(\omega)$ is
suppressed at 10 K shifts to lower frequencies as the CuO$_2$
layers in a unit cell decreases from 3 to 1.

\begin{figure}
\includegraphics[width=7.0cm,height=5.0cm]{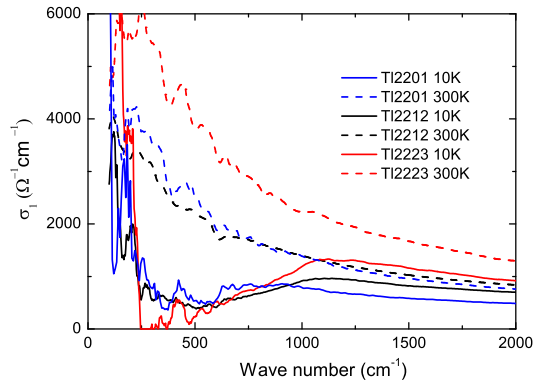}
\caption{\label{fig:epsart} (Color online) The low frequency
optical conductivity for the optimally doped Tl-2201, Tl-2212 and
Tl-2223 samples at 300 K and 10 K.}
\end{figure}

In fact, the change of optical response in the superconducting
state with respect to the normal state could be seen more clearly
from the plot of the ratio of the reflectance below $T_{c}$ over
that above $T_{c}$. Fig. 9 shows the ratio of
R$_{10K}$($\omega$)/R$_{130K}$($\omega$) as a function of
frequency. The R$_{10K}$($\omega$)/R$_{95K}$($\omega$) of a
Tl-2201 crystal and the R$_{10K}$($\omega$)/R$_{120K}$($\omega$)
of a Tl-2212 crystal are also included for comparison. Obviously
both of them have a maximum and a minimum in the plot. The values
near the maximum have much noise, nevertheless, we can see clearly
the dip minimum and the fitting maximum of Tl-2223 shift to higher
energy compared with Tl-2212\cite{Wang} and Tl-2201.

\begin{figure}
\includegraphics[width=7.0cm,height=5.0cm]{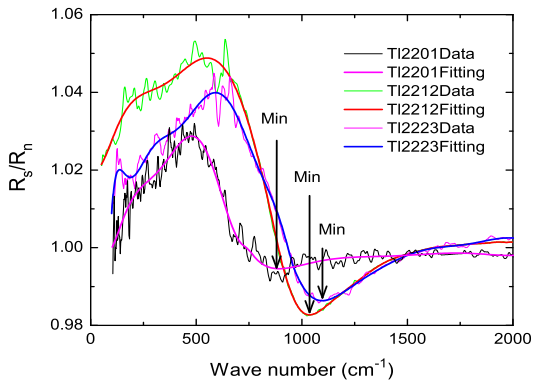}
\caption{\label{fig:epsart} (Color online) The
R$_{s}$($\omega$)/R$_{n}$($\omega$) from 100 to 2200 cm$^{-1}$ of
the optimally doped Tl-2223 single crystal with $T_{c}$=119K. The
Tl-2201 and Tl-2212 data have been included in for comparison. }
\end{figure}

A useful way to display in-plane infrared data is in terms of the
optical life time of the carriers which can be extracted from the
conductivity data using the extended-Drude formalism. The
reciprocal of the lifetime, or the scattering rate,
1/$\tau$(${\omega}$), is defined as\cite{Puchkov2}

\begin{eqnarray}
\tau^{-1}(\omega)=
 \frac{\omega_p^2}{4\pi}
 \frac{\sigma_1(\omega)}{\sigma_1^2(\omega)+\sigma_2^2(\omega)},
\end{eqnarray}
where $\omega$$_p$ is the plasma frequency. The scattering rate
1/$\tau$(${\omega}$) for 10 K and 130 K and 300 K data are shown
in Fig. 10, extracted from Eq.(1) using the plasma frequency
2.0$\times$10$^4$ cm$^{-1}$, determined by summarizing the optical
conductivity up to the plasma edge near 8,000 cm$^{-1}$. The
1/$\tau$(${\omega}$) at 300 K is linear in the low energy range.
However, 1/$\tau$(${\omega}$) at 10 K exhibits a rapid rise and a
substantial overshoot of the normal-state spectrum at the
frequencies corresponding to the maximum and minimum in the
R$_{10K}$($\omega$)/R$_{130K}$($\omega$) plot (in Fig.9),
respectively. Above this range the scattering rate in the
superconducting state becomes linear up to several thousand wave
number.

\begin{figure}
\includegraphics[width=7.0cm,height=5.0cm]{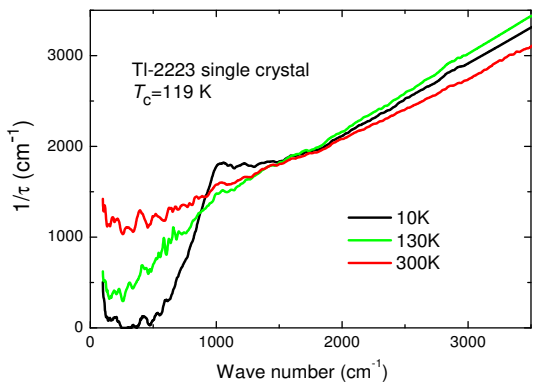}
\caption{\label{fig:epsart} (Color online) Optical scattering rate
spectra from 100 to 3500 ${cm^{-1}}$ of the optimally doped
Tl-2223 single crystal with $T_{c}$=119K.
 }
\end{figure}

The sharp rise and the overshoot in 1/$\tau$(${\omega}$) at low T
are quite similar to other high-T$_c$ cuprates. The difference is
that the energy scales of the characteristic features for Tl-2223
are higher than that for Tl-2212 compound. In our earlier
study\cite{Wang}, we already showed that the same structures in
1/$\tau$(${\omega}$) on Tl-2212 system with T$_c$=108 K are
substantially higher than systems with T$_c$ around 90 K. A
comparison of the scattering rate for the three phases at 10 K is
shown in Fig. 11. In this plot, the data for Tl-2201 crystal near
optimal doping are taken from Puchkov's work\cite{Puchkov1}. Our
data on Tl-2201 crystal are almost identical to their reported
data. The energy shifts of both the sharp rise and the overshoot
structures are seen rather clearly. In fact, the energy scales of
those features continue to shift up in compounds with higher
superconducting transition temperature. This can be seen very
clearly from optical work done on Hg-1223 with T$_c$$\sim$130
K\cite{McGuire}.

\begin{figure}
\includegraphics[width=7.0cm,height=5.0cm]{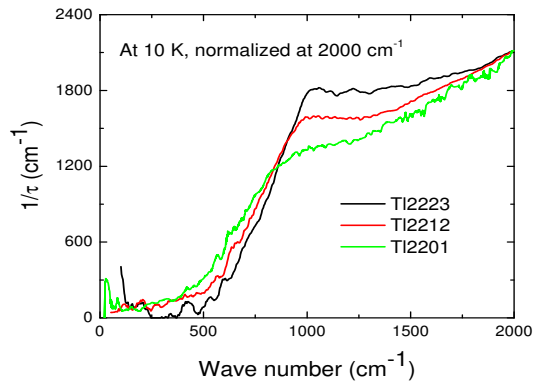}
\caption{\label{fig:epsart} (Color online) Optical scattering rate
spectra from 200 to 2000 ${cm^{-1}}$ of the optimally doped
Tl-2201, Tl-2212 and Tl-2223 single crystals. Data for Tl-2201
with T$_c$=88 K (very close to the optimal doping) are taken from
ref.\cite{Puchkov1}. The data have been normalized to Tl-2223 at
2000 cm${^{-1}}$.
 }
\end{figure}

\begin{figure}
\includegraphics[width=7.0cm,height=9.0cm]{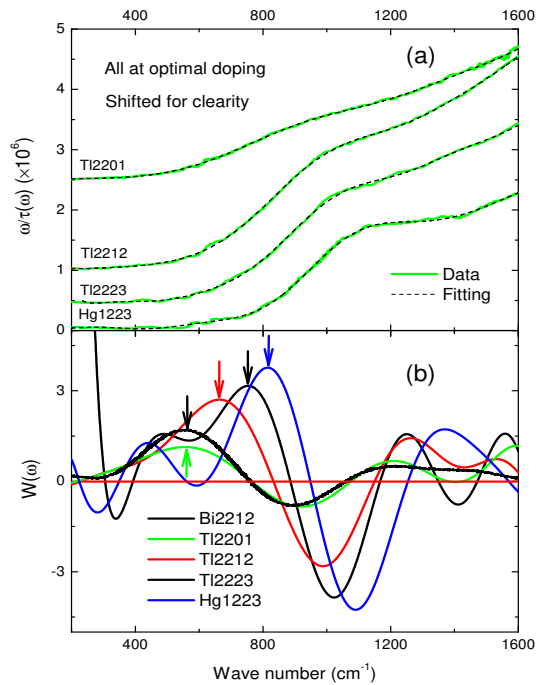}
\caption{\label{fig:epsart} (Color online) (a) Scattering rate
data multiplied by $\omega$ for Tl-2201, Tl-2212, Tl-2223, and
Hg-1223. The scattering data for Tl-2201 were taken from
\cite{Puchkov1}, the data for Hg-1223 from \cite{McGuire}. The
curves are shifted vertically for clarity. (b) The spectral
function W($\omega$) at 10 K (in the superconducting state)
derived from the polynomial fitting. The data for Bi-2212 were
taken from \cite{Tu}. The arrows indicate the positions of main
peaks.
 }
\end{figure}

The characteristic features were widely believed to result from
the interaction of electrons with a collective mode in combination
with a superconducting gap that appears below
$T_{c}$\cite{Carbotte1,Schachinger,Abanov,Tu,Hwang2}. In
photoemission experiment, a kink in the dispersion along the zone
diagonal develops in the low temperature which is due to the
correction by the real part of self-energy of quasiparticle, and
by Kramers-Kronig relation, a rapid broadening of the
photoemission line width, i.e. a sudden change of the life time of
quasiparticle is seen simultaneously\cite{Norman}. The enhanced
absorption above the knee frequency in infrared response results
from an enhanced scattering, which should have the same origin as
the line width broadening or the kink in the dispersion observed
along the zone diagonal in ARPES. Extracting the Bosonic spectral
function from the infrared spectra has been established
theoretically by performing second derivative of the optical
scattering rate multiplied by frequency
through\cite{Marsiglio,Carbotte1}:
\begin{equation}
   W(\omega)=
   {1\over2\pi}{d^2\over d\omega^2}[\omega{1\over\tau(\omega)}],
\label{chik}
\end{equation}
As such second derivative contains detectable singularities at a
number of characteristic energies, the inversion method was widely
used to determine the parameters like gap amplitude $\Delta$ and
Boson mode energy $\Omega$.

Fig. 12(a) shows the plot of the scattering rate multiplied by
frequency vs. frequency for Tl-2201, Tl-2212, and Tl-2223
crystals. The data for Hg-1223 crystal taken from
ref.\cite{McGuire} are also plotted. In order to perform second
derivative, the curves were fit with a high-order (40 orders)
polynomial. To see the energy scales of the features more clearly,
we have shifted the curves vertically for different compounds. The
spectral functions W($\omega$) obtained by equation (2) are
plotted in Fig. 12(b). The W($\omega$) data for a high-quality
Bi-2212 crystal with T$_c$=91 K obtained by Tu et al. are also
added\cite{Tu}. We can see clearly that, roughly at frequencies
corresponding to the sharp rise and the overshoot in
1/$\tau$(${\omega}$), there exist a major peak, or positive
maximum as indicated by the arrow and a negative minimum in
W($\omega$), respectively. In particular, for those systems with
different T$_c$ at optimal doping, the positive maximum and the
negative minimum increase systematically with T$_c$.

It also deserves to remark that some weak maxima could exist
preceding the main peak in W($\omega$). They come from the week
fluctuations in the fitting curves due to the experimental noise
in the far-infrared region. Because the signal-to-noise level at
low-$\omega$ side depends strongly on the sample size, those
structures are observed most eminently for Tl-2223 and Hg-1223
samples due to their relatively smaller sizes. Those maxima are
not steady for different measurements on different samples.
Nevertheless, the main peak and the following negative minimum in
each spectral function W($\omega$) are fairly robust. They
correspond to the sharp rise and overshoot in the scattering rate
spectra, or the knee structure and dip-like feature in the
reflectance spectra.

Although it is generally agreed that the second derivative of the
optical conductivity, i.e. the W($\omega)$ spectrum, reflects the
bosonic spectral function, different opinions exist regarding to
the energy positions of the positive maximum and the negative
minimum. Considering a d-wave gap symmetry, Carbotte et
al.\cite{Carbotte1,Schachinger} argued that the peak in
W(${\omega}$) below T$_{c}$ occurs at $\Delta$+$\Omega$, while the
minimum is at 2$\Delta$+$\Omega$. However, Abanov et
al.\cite{Abanov} argued that the d-wave gap does not affect the
threshold position in conductivity, as a result, a sharp maximum
in W($\omega$) occurs at 2$\Delta$+$\Omega$ followed by a deep
minimum. They emphasized that at T=0 K the maximum and minimum are
at the same frequency, while at finite temperature, the maximum
shifts to low frequency, but the minimum remains at the same
frequency as at T=0 K. So, they also identify the
2$\Delta$+$\Omega$ with the deep minimum in W($\omega$). If we
accept that the peak in W(${\omega}$) locates at
$\Delta$+$\Omega$, and the minimum is at 2$\Delta$+$\Omega$, we
obtained a plot for the peak and the dip energy as a function of
T$_c$ for the different systems at optimal doping, as shown in
Fig. 13. Obviously, both of them scale well with T$_c$ for
different systems. Furthermore, the difference between them should
give the superconducting gap energy $\Delta$. In the figure, we
also make the plot for the energy difference between the dip and
the peak vs T$_c$ for different compounds. It then shows a trend
of increase with decreasing the superconducting transition
temperature T$_c$ for different systems. This result is not
physically reasonable, and also in contradiction to the
ARPES\cite{Matsui,Feng} and Raman\cite{Gallais} experimental
results which show a decrease of the gap amplitude for the systems
with lower T$_c$. We think that the current analysis puts new
challenge to the established models.

\begin{figure}
\includegraphics[width=7.0cm,height=5.0cm]{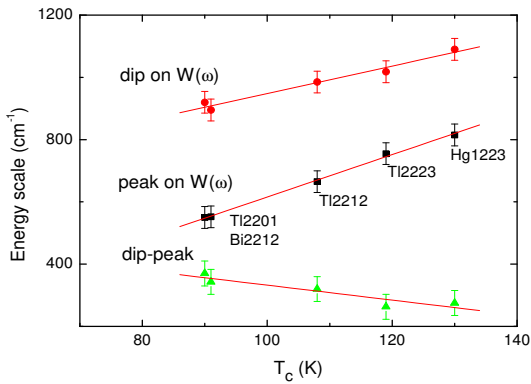}
\caption{\label{fig:epsart} (Color online) The energy scales of
the negative dip and the main peak vs. T$_{c}$ for several
optimally doped cuprates. The energy difference between the dip
and the peak is extracted simply.
 }
\end{figure}

Nevertheless, the linear scaling of the energy of the peak (and
also the dip) in W($\omega$) with T$_c$ is firmly established in
our experiment. At present, the nature of the mode is not clear.
It is still under dispute whether the mode is a phonon or a
magnetic resonance as detected by neutron scattering experiment.
If the maximum or minimum indeed involve a sum of the gap $\Delta$
and a Boson mode $\Omega$, then the scaling behavior not only
means that the gap amplitude $\Delta$ is proportional to T$_c$,
but Boson mode energy $\Omega$ is also proportional to T$_c$ for
the different systems. In this case, the phonon origin for the
Boson mode is unambiguously ruled out, because the in-plane phonon
is determined by the Cu-O bond length, which should not change in
systems with different T$_c$.

However, we would also like to point out that the above scenario
of a mode plus a gap for the reflectance feature below T$_c$
relies on the clean limit of the ab-plane superconductivity. The
unreasonable result obtained from the energy difference between
the dip and the peak in terms of the model by Carbotte et
al.\cite{Carbotte1} may suggest that the ab-plane of high-T$_c$
cuprates may not be in clean limit. In the case of dirty limit, on
the other hand, the sharp rise feature in the scattering rate
should be at 2$\Delta$, since the impurities elastic scattering
should be able to transfer the momentum of the particle-hole
excitations without affecting the threshold energy. At present,
there is still no agreement on whether or not the ab-plane is in
clean limit. Very recently, Homes\cite{Homes2} suggests that the
criteria of a small value of the quasiparticle scattering rate for
T $\ll$T$_{c}$ is not a good measure of whether or not the
superconductivity is in the clean or dirty limit. By contrast, The
normal-state value of 1/$\tau$ should be considered when
determining whether a system is in the clean or dirty limit. They
also pointed out that the clean-limit requirement is much more
stringent for a d-wave system than it is for a material with an
isotropic energy gap, since it not only requires 1/$\tau$ much
less than the maximum gap amplitude, but also
1/$\tau$$\leq$2$\Delta_{k}$ in the nodal regions. They established
a linear scaling relation $\rho_s\propto\sigma_{dc}T_c$ for
cuprates\cite{Homes3}, and argued that this scaling is the
hallmark of a dirty-limit system\cite{Homes2}. In a dirty-limit
case, no Boson mode would be involved in the sharp rise of
scattering rate, or the main peak in the spectral function
W($\omega$). The feature would be purely due to the singularity of
density of state at gap position. Then what we observed and
analyzed in this work only reflect the variation of the gap
amplitude in systems with different T$_c$. At present, it needs
rather sufficient experimental data and more theoretical efforts
to identify whether the ab-plane superconductivity in cuprates is
in the dirty-limit.

\section{Summary}

Single crystals of thallium-based cuprates have been grown by flux
method. All crystals were of the platelet form, from 0.5 to
2mm$^2$. They were characterized by ac susceptibility, X-ray
diffraction, SEM and EDX analysis. The optical properties were
investigated for the crystals with focus on the Tl-2223 compounds.
The reflectance spectrum exhibits a knee structure at around 600
cm$^{-1}$ and a dip at higher frequency below $T_{c}$. The ratio
of the reflectance below and above Tc displays a maximum and
pronounced minimum at the knee and dip frequencies. The scattering
rate and its second derivative were extracted and compared with
other cuprate systems. A linear scaling behavior for the maximum
and the negative minimum with T$_c$ is established for systems
with different T$_c$ at optimal doping. We discussed the physical
consequences of the experimental data in terms of both clean and
dirty limits. At present, it is an open question whether the
ab-plane of HTSCs is in the clean or dirty limit.

\begin{acknowledgments}
We wish to acknowledge senior engineer H. Chen for her help in XRD
experiment. This work is supported by National Science Foundation
of China, the Knowledge Innovation Project of Chinese Academy of Sciences.
\end{acknowledgments}







\end{document}